\documentclass[12pt,fleqn]{article}

\usepackage{graphicx}
\usepackage{pdftricks}
\usepackage{lineno}
\usepackage{dcolumn}

\begin{document}

\title{\Large Some Early Ideas on the Metric Geometry of Thermodynamics}

\author{
   George Ruppeiner\footnote{ruppeiner@ncf.edu}\\
   Division of Natural Sciences\\
   New College of Florida\\
   5800 Bay Shore Road\\
   Sarasota, Florida 34243-2109}

\maketitle

\begin{abstract}

It is a pleasure to write for this 90'th anniversary volume of Journal of Low Temperature Physics dedicated to Horst Meyer at Duke University.  I was a PhD student with Horst in the period 1975-1980, working in experimental low temperature physics.  While in Horst's group, I also did a theoretical physics project on the side.  This project in the metric geometry of thermodynamics was motivated by my work in Horst's lab, and helped me to understand the theory of critical phenomena, very much in play in Horst's lab.  In this paper, I explain the essence of my theory project and give a few accounts of it's future development, focussing on topics where I interacted with Horst.  I pay particular attention to the pure fluid critical point.

\end{abstract}


\section{Introduction}

It was a privilege for me to have worked in Horst Meyer's group at Duke University from the period 1975-1980.  I received a PhD in experimental low temperature physics under Horst's directions in 1980 \cite{Ruppeiner1980}.  My project had me measure transport properties in $^3$He-$^4$He fluid mixtures near the tricritical point at temperatures around $1$ K.  Although after graduation I did not continue to work in experimental low temperature physics, what I learned working in Horst's group completely shaped my future research direction.

\par
In this 90'th anniversary issue of Journal of Low Temperature Physics (JLTP) dedicated to Horst, I take the opportunity to discuss theoretical work I began on the side at Duke in the area of metric geometry of thermodynamics.  This theoretical contribution augmented the structure of basic thermodynamic fluctuation theory with an arguably superior one.  It extends the reach of basic macroscopic thermodynamics into mesoscopic size scales, where many important thermodynamic properties are determined.

\par
Thinking in unorthodox ways can sometimes lead to improvements in even basic theory.  But new ideas always emerge from a context.  For my theoretical work, the colleagues, the research topics, the financial support, and the general working environment in Horst's experimental group were absolutely essential.

\section{My Start at Duke}

Shortly after I was admitted to Duke University as a Physics graduate student, Horst invited me to work in his group during the summer of 1975.  These were exciting days in Horst's lab, with much talk about the recent (1972) discovery of the superfluid phases in $^3$He at millikelvin temperatures at Cornell University.\footnote{Robert Richardson, one of Horst's former PhD students, shared a Nobel Prize for this discovery.}  I remember especially hearing first-hand about this find from Horst's postdoc Moses Chan from Cornell.

\par
That summer, I measured sound velocity and attenuation near the tricritical point under the supervision of graduate student David Roe.  This project led to my first publication \cite{Ruppeiner1977}, and started me reading about the theory of critical phenomena.  But my background in statistical mechanics and thermodynamics was relatively weak, so I found these readings rather challenging.

\par
However, I could readily grasp the agenda for experimentalists set down by the emerging ideas of scaling and universality.  Particularly clear were the ideas of Widom in 1965 \cite{Widom1965}, who proposed that the picture from the ''classical'' van der Waals equation, with its power law divergences and universal scaled equation of state, could be generalized to real pure fluid systems by employing ''non-classical'' critical exponents and scaled equation of state.  Such new ideas were necessary since by the mid-1960's measured deviations between the critical properties of van der Waals and real fluid systems had become too large to ignore.

\par
But I found the underlying microscopic justification for scaling and universality difficult to follow in any detail.  Discussion was usually given in terms of the newly developed (1971) renormalization group theory (RG), employed in the context of many standard models of statistical mechanics.  Mathematically, RG is very difficult, with few cases leading to exact solutions.  In addition, I found the field theory language employed in this theory daunting.\footnote{But this language is very powerful.  If we must understand macroscopic physics by building up from the level of the molecules, there is simply no rational alternative.  If I had been much more sophisticated theoretically, and had access to a book such as the one by Altland and Simon \cite{Altland2010}, my future direction in theory might have been quite different.}

\par
Definitely helpful were Horst's regular communications with a number of theorists adept at communicating their ideas in the language of the experimentalist.  I soon discovered that such theorists were in somewhat short supply, but Horst knew how to find them.  Horst organized an active seminar series, with theoretical speakers always advised to emphasize measurable ideas.  Despite this effort, the students in Horst's lab were always conscious of the fact that there was a great divide between experiment and theory, one hard to bridge.  Personally, I was always comfortable with the direct, clear language of the low temperature experimental physicist.  I found theorists tougher to follow.  They tended to divide into various camps whose relation to experiment could be hard to sort out.

\section{My First Year of Graduate Classes}

Towards the conclusion of my first summer at Duke, I began registration for my fall graduate classes.  My schedule had a slot for a Mathematics class, and I selected General Relativity, taught by Murray Cantor in the Mathematics department.  I had long been interested in this subject because as an undergraduate student at LSU my advisor was William Hamilton, who was building the first low temperature gravitational wave detector at LSU.  This detector was a precursor to LIGO.\footnote{In 1971, I took Physics I as a freshman with Bill, and he promised us students gravity wave data when we were sophomores.  This timetable turned out to be a bit too optimistic (finally, LIGO 2016!), but I remember being very impressed by the spirit of the thing.}  My interest in the coming General Relativity course was further boosted by the appearance in the Duke bookstore of the just published (1973) textbook: the spectacular {\it Gravitation}, by Misner, Thorne, and Wheeler \cite{Misner1973}.

\par
In my first semester of graduate school, I spent most of my time in class, with some data taking in Horst's lab (this was expected).  I found myself mostly well prepared to start graduate classes, especially in quantum mechanics, which Ravi Rau at LSU taught me.  But particularly interesting to me was the General Relativity course.  The fact that Einstein set down a theory (1915) which continued to dominate all its competition, which was readily accessible in textbooks, and which allowed the direct calculation of a number of important results, was intellectually overwhelming.

\par
My second semester brought a key event for me: the publication in March of 1976 of a paper in Physics Today by Frank Weinhold, who presented a metric geometry of thermodynamics \cite{Weinhold1975}.\footnote{Michael Ryschkewitsch, a graduate student in Horst's lab, showed me this paper when it was published.  It was natural that this paper would be shown to me, since my lugging {\it Gravitation} around the lab marked me as having a bit of a ''mathematical bent,'' as Horst put it.}  Weinhold introduced a positive-definite inner product between thermodynamic ''vectors'' that simplified calculations involving thermodynamic response functions, and gave such calculations a geometric meaning.  When Weinhold's paper was published, any metric structure in thermodynamics was novel, since prevailing opinion was that thermodynamics allowed no such structure.

\par
I found Weinhold's geometric structure to be very appealing.  Given my study of the geometry in general relativity, I thought that  Weinhold's geometry might offer some insight into critical phenomena.  This idea really had no rational basis, it is just that people like to attack difficult problems with the tools at their disposal, and my set of tools was rather limited.

\section{Metric Thermodynamic Geometry}

My initial thoughts about connecting Weinhold's geometry to critical phenomena were not very productive, and I concluded that new ideas were needed.  Weinhold's metric really offered no clear physically motivated rule for distance between thermodynamic states.  Weinhold's metric is really just an inner product, intended only to represent the second law of thermodynamics.  But a physically motivated distance rule is essential for a full differential metric geometry, including an induced curvature of a type that is so productive in general relativity.

\par
At this point, study of the Landau and Lifshitz statistical mechanics book \cite{Landau1977} gave me an idea.  These authors devote their Chapter XII to the topic of thermodynamic fluctuation theory.  Although Landau and Lifshitz did not explicitly say so, fluctuation theory offers a natural probability metric.\footnote{Thermodynamic fluctuations are usually presented in the books as an add-on to thermodynamics, and not really necessary.  But I remember being very impressed at the time by an old paper by G. N. Lewis arguing that, in fact, fluctuation theory was logically necessary for thermodynamics \cite{Lewis1931}.}

\par
Let me describe this probability metric.  Consider some very large thermodynamic pure fluid system $\mathcal{A}_0$ in equilibrium, and with fixed temperature and density $T_0$ and $\rho_0$, respectively.  At large size scales, $\mathcal{A}_0$ looks smooth and uniform, but at progressively smaller sizes we find increasing ''jitter,'' as the more or less continual random motion of the molecules becomes increasingly evident.  A somewhat remarkable find by Einstein (1907) was that this jitter could be quantified with thermodynamics.

\par
\begin{figure}
\begin{minipage}[b]{0.5\linewidth}
\includegraphics[width=2.7in]{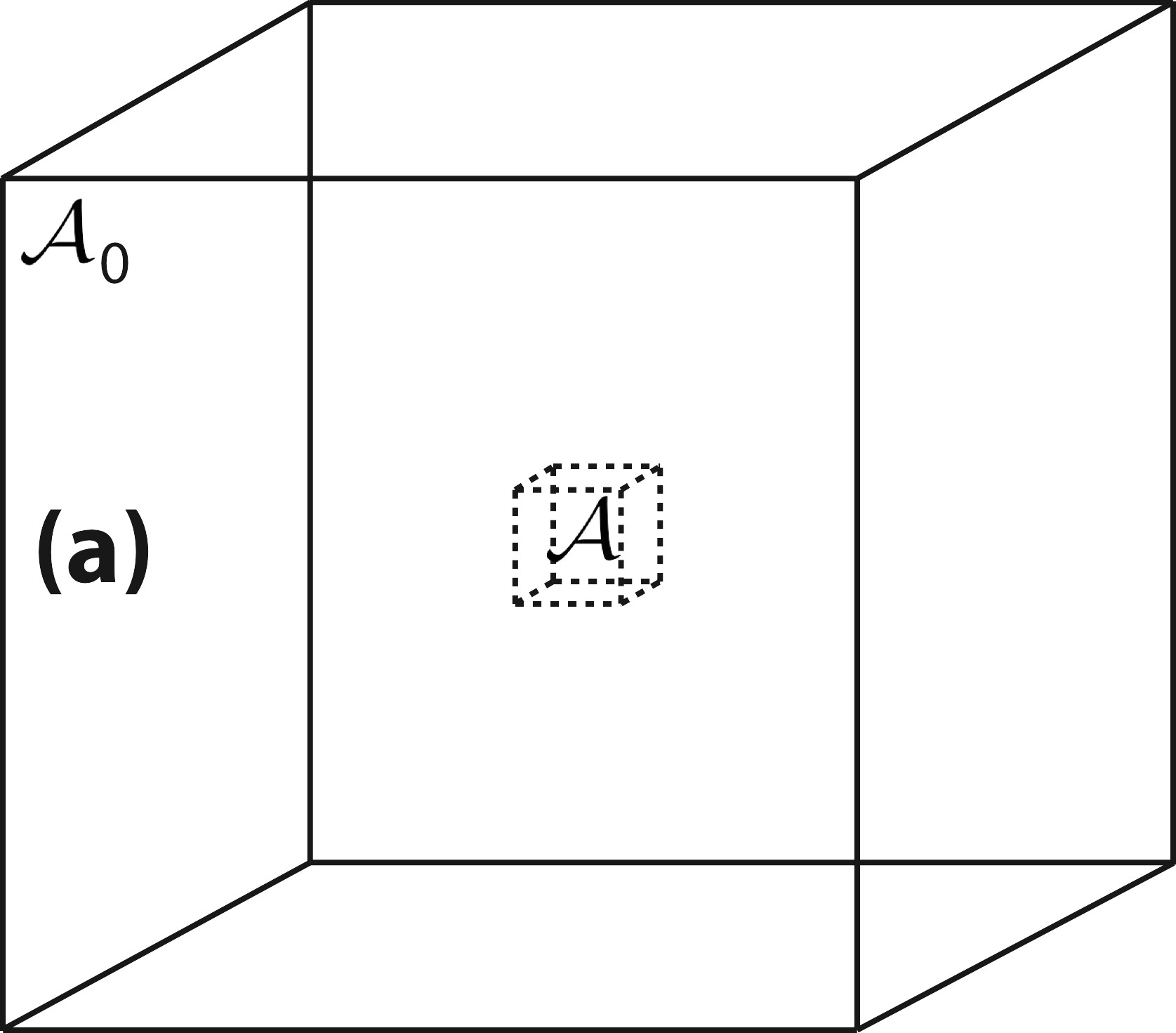}
\end{minipage}
\hspace{0.0 cm}
\begin{minipage}[b]{0.5\linewidth}
\includegraphics[width=2.7in]{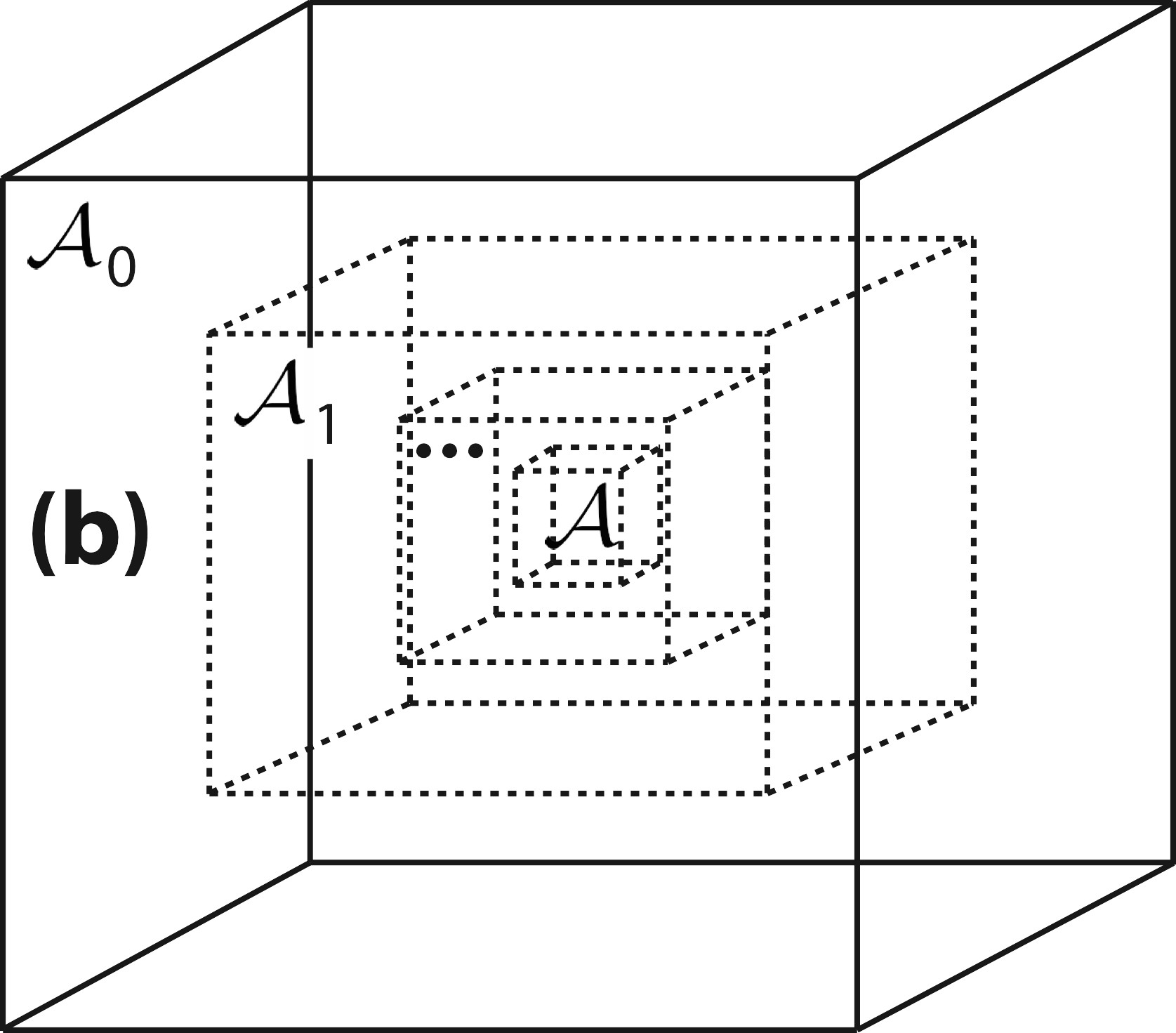}
\end{minipage}
\caption{Two basic theoretical structures: a) the standard structure of Landau and Lifshitz \cite{Landau1977}, showing a single open subsystem $\mathcal{A}$ of an infinite environment $\mathcal{A}_0$, and b) an extended structure which I first proposed showing a hierarchy of $n+1$ concentric open subsystems of decreasing size: $\mathcal{A}_1$, $\mathcal{A}_2$, $\cdots$, $\mathcal{A}_n$, $\mathcal{A}$, with each subsystem sampling only the state of the subsystem immediately larger than it.  The goal in both cases is to find the fluctuation probability of $\mathcal{A}$.}
\label{fig:1}
\end{figure}

To see how this is done, consider an open subsystem $\mathcal{A}$, with constant volume $V$, of $\mathcal{A}_0$.  $\mathcal{A}$ has fluctuating temperature and density $(T,\rho)$.  This structure is shown in Figure \ref{fig:1}a.  According to the familiar Gaussian form of the theory \cite{Landau1977}, the probability density $P$ for a fluctuation of the state of $\mathcal{A}$ to a temperature and density in a small neighborhood of $(T,\rho)$ is

\begin{equation}P\propto\mbox{exp}\left(-\frac{V}{2}\Delta\ell ^2\right),\label{10}\end{equation}

\noindent where

\begin{equation}\Delta\ell^2=\frac{1}{k_B T}\left(\frac{\partial s}{\partial T}\right)_{\rho}\Delta T^2+\frac{1}{k_B T}\left(\frac{\partial\mu}{\partial\rho}\right)_T\Delta\rho^2,\label{20}\end{equation}

\noindent $\Delta T=T-T_0$, and $\Delta\rho=\rho-\rho_0$.  Also, $k_B$ is Boltzmann's constant, $s$ is the entropy per volume, and $\mu$ is the chemical potential.  The coefficients of 
 $\Delta T^2$ and $\Delta\rho^2$ in Eq. (\ref{20}) are evaluated in the state $(T_0,\rho_0)$.
 
\par
Clearly, the quadratic form $\Delta\ell^2$ looks formally very much like a distance between points in the geometry of curved surfaces.\footnote{I use the intuitive word ''surface'' rather than the mathematically more accurate ''manifold.''  ''Surface'' implies being embedded in a 3D flat space, not intended here.  But for the purposes of visualization, such technicalities will not concern us in this paper.}  Notice that it is never negative, by thermodynamic stability.  A physical meaning for the distance $\Delta\ell^2$ between thermodynamic states is evident from Eq. (\ref{10}): the less the probability of a fluctuation between two states, the further apart they are.\footnote{When I first got this idea for probability distance, I thought that it was a novel find since I had seen it nowhere in the statistical mechanics literature.  However, probability distance is an element of information theory in the form of the Fisher information metric, which dates back well before my efforts.  My contribution was to help bring the idea of probability distance to thermodynamics, and then, in particular, to pay serious attention to the induced thermodynamic curvature $R$.}

\par
A quadratic line element, or metric, as in Eq. (\ref{20}) is commonly encountered in any undergraduate physics course involving surfaces, particularly courses using coordinates other than Cartesian coordinates.  For a flat surface in Cartesian coordinates $(x,y)$, the line element takes the simple form:

\begin{equation}\Delta\ell^2=\Delta x^2+\Delta y^2.\label{30}\end{equation}

\noindent We may alternatively employ polar coordinates $(r,\theta)$ for the plane:

\begin{equation}\Delta\ell^2=\Delta r^2+r^2\Delta \theta^2,\label{40}\end{equation}

\noindent where $r$ is the distance from the origin and $\theta$ is the polar angle.

\par
An essential element in the metric geometry of surfaces is that fundamental properties like the length of curves or the angles between vectors (also given by the line element) are independent of the coordinates used to calculate them.  Length and angle are said to be invariant quantities, the same calculated with either line element Eq. (\ref{30}) or (\ref{40}), for example.

\par
Of course not all surfaces are flat like the plane.  For example, a sphere with radius $a$, and parameterized by the polar and azimuthal angles, $\theta$ and $\phi$, respectively, has line element:

\begin{equation}\Delta\ell^2=a^2\Delta\theta^2+a^2 \mbox{sin}^2\theta\,\Delta\phi^2.\label{50}\end{equation}

\noindent Generally, a curved surface has a local geometry that differs fundamentally from that of a flat geometry.  For example, the circumference $C$ of a small circle of radius $r$ drawn on a curved surface is given by

\begin{equation} C=2\pi r+\frac{\pi }{6}R\,r^3+O(r^4),\label{60}\end{equation}

\noindent where the scalar curvature $R=0$ for the plane and $R=-2/a^2$ for the sphere.

\par
$R$ gives the answer to all questions regarding the deviation of local geometry from Euclidean geometry.  $R$ may be calculated from the line element.  For example, with a diagonal line element in $(x_1,x_2)$ coordinates:

\begin{equation}\Delta\ell^2=g_{11}\,\Delta x_1^2+g_{22}\,\Delta x_2^2,\label{70}\end{equation}

\noindent the curvature is \cite{Ruppeiner2010}

\begin{equation}R=\frac{1}{\sqrt{g}}\left[\frac{\partial}{\partial x_1}\left(\frac{1}{\sqrt{g}}\frac{\partial g_{22}}{\partial x_1}\right) + \frac{\partial}{\partial x_2}\left(\frac{1}{\sqrt{g}}\frac{\partial g_{11}}{\partial x_2}\right)\right],\label{80}\end{equation}

\noindent where $g$ is the product of the metric elements: $g=g_{11}\,g_{22}$.

\par
$R$ is an invariant.  Namely, although its value may vary from point to point on the surface, its value at a given point is the same regardless the coordinate system used to calculate it.  For example, for the flat plane $R=0$ computed in either Cartesian or polar coordinates.

\section{Physical Interpretation of the Thermodynamic $R$}

These ideas from metric surface geometry find immediate thermodynamic application with the introduction of the thermodynamic metric in Eq. (\ref{20}).  For example, the thermodynamic curvature $R$ is immediately computable from Eqs. (\ref{20}), (\ref{70}), and (\ref{80}).  Since $R$ is one of the few invariants in this thermodynamic geometry, one might expect it to be telling us something pretty important.  But what?  A priori, this is not at all clear.

\par
To find out, I chose simply to calculate $R$ for some known cases.  Start with the simple ideal gas, with Helmholtz free energy \cite{Landau1977}

\begin{equation} f(T,\rho)=\rho k_B T \ln\,\rho+\rho k_B h(T),\label{90}\end{equation}
 
\noindent where $h(T)$ is some function of the temperature $T$ with negative second derivative.  $h(T)$ corresponds to various internal properties of the constituent molecules.  We have $\{s,\mu\}=\{-f_{,T},f_{,\rho}\}$, where the comma notation indicates differentiation.

\par
By Eqs. (\ref{20}) and (\ref{90}), the thermodynamic line element is

\begin{equation}\Delta\ell^2=-\frac{\rho h''(T)}{T}\Delta T^2 + \frac{1}{\rho}\Delta\rho^2.\label{100}\end{equation}

\noindent Eq. (\ref{80}) now yields for the ideal gas:

\begin{equation}R=0,\label{110}\end{equation}

\noindent regardless of the function $h(T)$.  This is a very pleasing result, and it led me to speculate that $R$ might be some sort of measure of the interactions between molecules, absent in the ideal gas.  The fact that the intrinsic properties of the constituent molecules, embedded in the function $h(T)$, do not matter reinforces such an interpretation.

\par
But this result does not take us very far by itself.  I also noted that $R$ has units of volume, posing the question: what thermodynamic quantity having units of volume could conceivably tell us something about interactions between molecules?  This turned out to be a surprisingly easy question for me to answer given what I was learning about critical phenomena in Horst's group.

\par
I remember in particular reading a paper by Widom \cite{Widom1974},\footnote{I always found Widom's papers to be relatively nontechnical, free of heavy math and jargon, clear to read, and filled with good physical ideas.} who wrote about the role of the correlation length $\xi$ near the critical point.  In the statistical mechanics books, $\xi$ gives the range of the exponentially decaying pair correlation function.  This picture makes $\xi$ look like a microscopic quantity.  But Widom argued that $\xi^3$ was proportional to the inverse of the singular part of the free energy per volume $\phi$ (in units of $k_B$):

\begin{equation}\phi\propto\frac{1}{\xi^3}.\label{120}\end{equation}

\noindent Such a picture makes $\xi$ look like a macroscopic thermodynamic quantity.  Since $\phi$ is the source of all of the thermodynamics, the relation Eq. (\ref{120}) is rather important.  Indeed, Goodstein \cite{Goodstein1975} went so far as to say: ''Nothing matters except $\xi$.''\footnote{I add that Eq. (\ref{120}) leads to the hyperscaling critical exponent relation.}

\par
Inspired by these powerful ideas from critical phenomena, I made the somewhat obvious conjecture matching quantities with the same units:

\begin{equation}|R|\propto\xi^3\label{130}.\end{equation}

\noindent This certainly works for the ideal gas, with $R$ and $\xi$ both zero.  In addition, it is plausible near the critical point since it was found that both $|R|$ and $\xi$ diverge there.  In the picture of Eq. (\ref{130}), $\xi$ looks like a mesoscopic quantity, one given by thermodynamic fluctuation theory by means of $R$.

\par
However, a more detailed comparison was necessary to clinch the proportionality between $|R|$ and $\xi^3$ near the critical point.  Nowadays, such a task would be easier since there is ready access to a number of scaled equations of state.  But this was not the case when I was first doing these calculations.  Several papers I was reading at the time involved Ho and Lister's linear model for the full thermodynamic properties.  This model allowed me to work out $R$ along the critical isochores in five fluids: $^4$He, Xe, $^3$He, H$_2$O, and O$_2$.  (The $^3$He data were measured in Horst's lab  prior to my coming to Duke \cite{Wallace1970}.)

\par
Comparison of $R$ with experimentally measured values of $\xi$ then showed excellent agreement between $R$ and $\xi^3$, including both critical exponents and critical amplitudes.  I found

\begin{equation}\xi^3=-\frac{1}{2}R\label{140},\end{equation}

\noindent an asymptotic critical point equality that I subsequently verified in a number of other fluid and spin systems, both on and off the critical isochore, and in dimensions other than three (with the exponent $3$ in $\xi^3$ replaced by the spatial dimensionality $d$).

\par
While I was working on this theory, I was actually spending most of my time working on my experimental thesis project with Horst.  Nevertheless, I had now reached a point where my theoretical results could be written up in a manuscript.  I did this, with assistance from Horst, Richard Palmer, and Brian Buck (visiting from Oxford).  Baird Straughan, an undergraduate literature student, aided me considerably by carefully editing several drafts of my manuscript for good grammar.  I submitted my manuscript to Physical Review A, and they published it \cite{Ruppeiner1979}.  Horst very kindly took care of the page charges.

\par
I am sometimes asked why Landau and Lifshitz \cite{Landau1977} did not report the significance of the thermodynamic $R$.  Maybe, the reason is that they did not see it.  Notice, for example, that the ideal gas line element Eq. (\ref{100}) looks like neither of the flat space line elements, Eqs. (\ref{30}) or (\ref{40}).  A little work is required to bring it to one of these forms.  If the ideal gas result $R=0$ were evident at a glance from the line element, then I think that the interaction interpretation of the thermodynamic $R$ would have been developed long before I started my work in Horst's group.

\section{Covariant Fluctuation Theory}

The results described above might appear to have a somewhat mysterious quality to them: why should ideas from Einstein's general theory of relativity have anything to say about critical phenomena?  But I soon learned that there is nothing ''spooky'' about any of this.  The proportionality between $R$ and $\xi^3$ follows from extending the basic theoretical structure in thermodynamic fluctuation theory from that in Fig. \ref{fig:1}a to that in Figure \ref{fig:1}b.

\par
In this extended theoretical structure, instead of the single open subsystem $\mathcal{A}$ of the infinite system $\mathcal{A}_0$, we envision a hierarchy of $n+1$ concentric subsystems of $\mathcal{A}_0$ of decreasing size: $\mathcal{A}_1$, $\mathcal{A}_2$, $\cdots$, $\mathcal{A}_n$, $\mathcal{A}$ with $n\to\infty$, and with the difference in size between adjacent subsystems in the hierarchy going to zero.  The largest subsystem $\mathcal{A}_1$ is big enough that the fluctuations in its temperature and density are effectively zero.

\par
The basic motivation behind this hierarchical structure is to address a difficult problem: when $\xi$ gets large, correlated fluctuating groups of many molecules have formed, and these groups generally correspond to local thermodynamic states deviating widely from that of $\mathcal{A}_0$.  Hence, a subsystem $\mathcal{A}$ with volume $V\sim\xi^3$, or less, is likely to find itself lost in a local environment with state deviating widely from that of the overall average set by $\mathcal{A}_0$, and the structure in Fig. \ref{fig:1}a must be inadequate.  But with the hierarchical structure in Fig. \ref{fig:1}b, each subsystem samples only the state of the subsystem immediately larger than it, and there is never any substantial loss of contact with its environment.

\par
For this picture to be effective, a mathematical structure for calculating with it is required.  Such mathematics is offered by a path integral expression developed in the mid-1970s \cite{Grabert1979} for time-dependent irreversible thermodynamic processes corresponding to curved metrics.\footnote{I first learned about path integrals from reading Tisza's book \cite{Tisza1966} about Markov processes.  Also very interesting to me was Feynman's book on quantum mechanics and path integrals \cite{Feynman1965}, which Moses Chan gave me.}  Formally, this is the solution to the Fokker-Planck equation.  Although my physical application is quite different (there is no time dependence), the path integral formalism applies immediately to the hierarchical structure in Fig. \ref{fig:1}b if we replace time in the irreversible thermodynamic formalism by inverse volume.  The probability that $\mathcal{A}$ is found in some given state is now the sum of the contributions of all the possible ways the states of $\mathcal{A}_1$, $\mathcal{A}_2$, $\cdots$, $\mathcal{A}_n$ could be arranged.

\par
The result is a covariant and consistent thermodynamic fluctuation theory \cite{Ruppeiner1983a, Ruppeiner1983b, Diosi1985}.  In this theory, $|R|$ marks the volume scale where the fluctuations transition from the large volume Gaussian expression in Eq. (\ref{10}) to something different.  Physically, this volume ought to be $\xi^3$, establishing the proportionality between $|R|$ and $\xi^3$.

\section{Towards Graduation from Duke}
 
Given my engagement with a theory project, one might think that I should have pursued it as my PhD project, or even that my efforts in experimental physics were somehow misdirected.  But I liked my experimental project, worked hard at it, and I liked working with Horst and being a part of his group.  I never considered changing thesis projects.  Furthermore, my ideas in theoretical physics were motivated by the physical ideas in play in Horst's group.  Had I gone into theoretical physics, my focus would probably have been quite different.

\par
In addition, Duke University supported me by allowing me to take further courses in the Duke Math Department, with William Allard and Michael Reed.  As I became more experienced, I was promoted to Instructor (1980), on Horst's recommendation.  This paid pretty well and strengthened my resume.  All in all, I really would not have done anything different.

\section{Future Developments}

In the years since my graduation from Duke, a number of new results by a number of authors have emerged.  I go over some of them here, emphasizing my future interactions with Horst.

\subsection{{\it The Sign of $R$}}

\begin{figure}
\centering
\includegraphics[height=4cm]{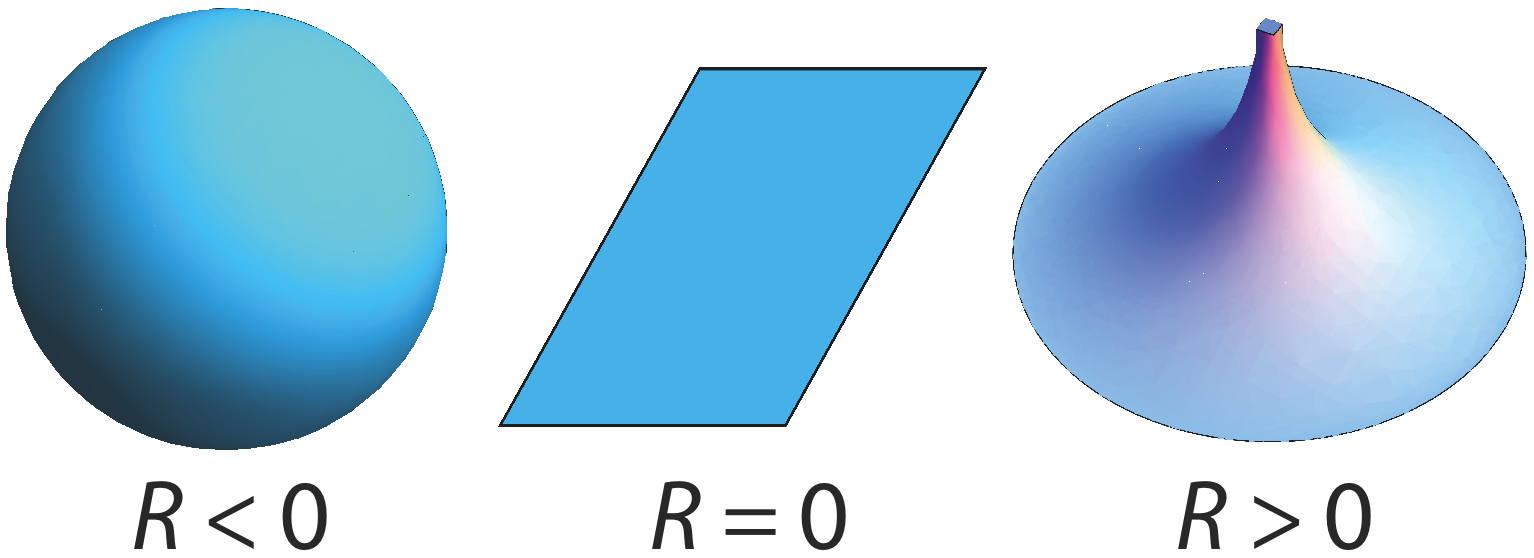}
\caption{Three surfaces with constant scalar curvature $R$: the sphere, the plane, and the pseudosphere.  Attractive interactions correspond to the geometry of the sphere, and repulsive interactions to the geometry of the pseudosphere.}
\label{fig:2}
\end{figure}

The sign of $R$ is important.  I did not originally pay much attention to it, since all of the cases that I had worked out at Duke (mostly critical points) had negative $R$.  However, Janyszek and Mruga{\l}a \cite{Mrugala1990, Oshima1999} found that the sign of $R$ is uniformly negative for the ideal Bose gas (in my curvature sign convention \cite{Weinberg1972}), and uniformly positive for the ideal Fermi gas.  This suggests that systems where attractive intermolecular potentials dominate (e.g., the familiar critical point models) have negative $R$, and systems where repulsive intermolecular potentials dominate have positive $R$.  This is the case even in the ideal quantum gases, where the interactions are purely statistical.  Figure \ref{fig:2} shows three characteristic surfaces with constant $R$.  I do not really know why thermodynamics should pick up on the character of the interactions in this way, but it seems to be a persistent feature \cite{Ruppeiner2010, Ruppeiner2014}.

\par
The sign of $R$ thus offers a classification of thermodynamic systems according to the character of their underlying intermolecular interactions.  One might say that such a classification could tell us little new, since if we just work out $R$ from the statistical mechanics of known models, then we will already know the interactions ahead of time.  While this may be so, there are significant cases where we know the thermodynamics, but do not know how to get it from any underlying microscopic model.  Either we cannot evaluate the partition function from the known microscopic Hamiltonian, or we do not know the microscopic Hamiltonian at all.  In such cases, the thermodynamic calculation of $R$ offers real hope of telling us something new, as I argue in the next two subsections.

\subsection{{\it $R$-Diagrams of Real Pure Fluids}}

The pure fluid is an example of a system where a lot is known about the intermolecular potentials, but where the actual evaluation of the partition function in all but the simplest cases is very difficult.  On the other hand, experimentally measuring fluid thermodynamic properties, and then fitting the resulting data to multiparameter functions for the Helmholtz free energy is straightforward.  This has been done for many fluids \cite{NIST}, and the calculation of $R$ is straightforward \cite{Ruppeiner2012b, Ruppeiner2015a}.

\par
\begin{figure}
\begin{minipage}[b]{0.5\linewidth}
\includegraphics[width=2.7in]{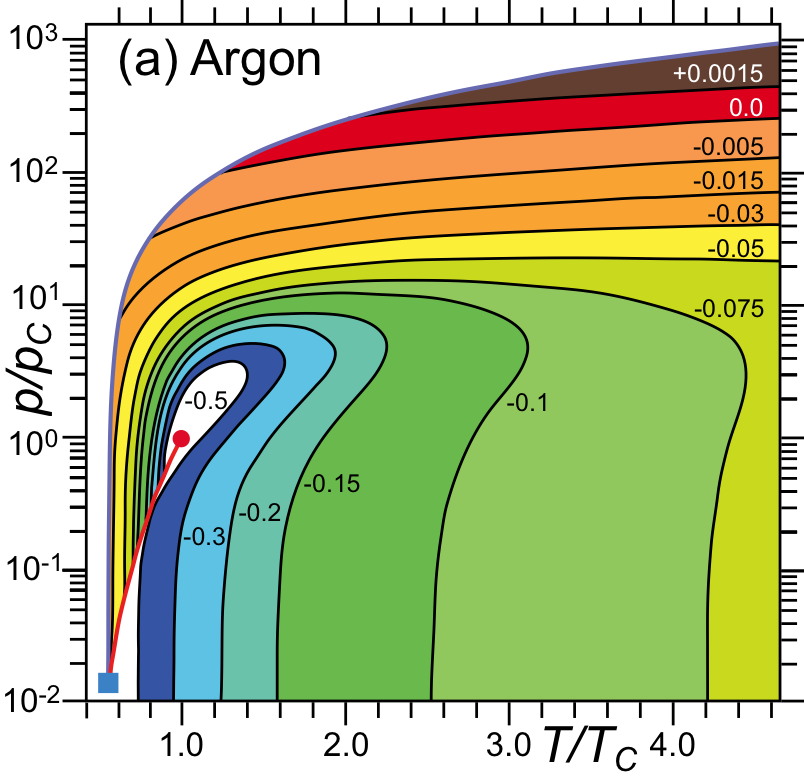}
\end{minipage}
\hspace{0.0 cm}
\begin{minipage}[b]{0.5\linewidth}
\includegraphics[width=2.7in]{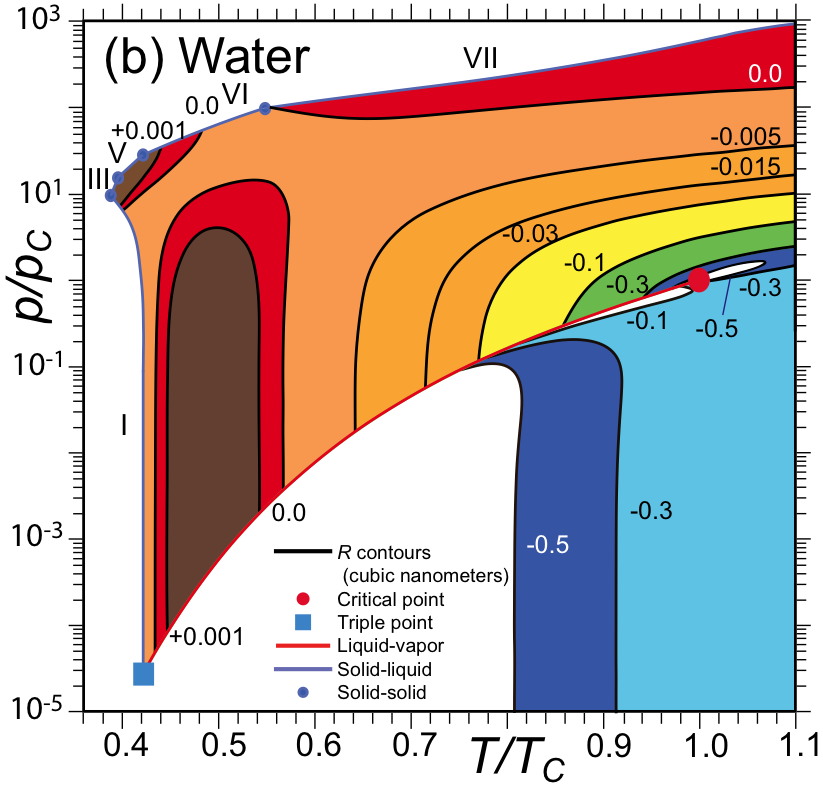}
\end{minipage}
 \caption{$R$-diagrams for a) argon, and b) water.  The units of the $R$-contours are cubic nanometers, and I plot temperature $T$ and pressure $p$ in units of their critical point values $T_c$ and $p_c$. Included in each plot are the triple point, the critical point, and the liquid-vapor and solid-liquid coexistence curves. Various solid phases in water are labeled.}
\label{fig:3}
\end{figure}

Figure \ref{fig:3} shows the $R$-diagrams for argon and water \cite{Ruppeiner2015a}.  Most of the fluid states have negative $R$, as the attractive tail of the intermolecular potential dominates over most of the phase diagram.  At larger densities, we might expect a transition to positive $R$ because solid phases are generally modeled to leading order with the hard-core repulsive part of the intermolecular potentials.  Indeed, positive $R$'s, with values on the order of cubic Angstroms, are present in both argon and water at very high pressure and temperature; see Fig. \ref{fig:3}.  Such regimes of positive $R$ were found in all the fluids that we looked at \cite{Ruppeiner2015a}.

\par
But water has an additional feature with positive $R$ in the neighborhood of ambient liquid water; see Figure \ref{fig:3}b ($T/T_c\sim0.5$).  $R$ peaks at about $+2$ cubic Angstroms in this regime.  Perhaps, this thermodynamic feature relates to a long-standing question about the microscopic structure of liquid water.  There is reason to think that liquid water could contain microscopic ''solid-like'' features consisting of open tetrahedral networks linked by hydrogen bonds \cite{Debenedetti2003}.  Such networks would explain a number of well-known thermodynamic anomalies in water, such as the density maximum at ambient $4^{\circ}$C.  However, direct microscopic evidence for such molecular networks is hard to come by, and the thermodynamic connection in \cite{Debenedetti2003} is somewhat indirect.  Perhaps, the ''solid-like'' positive $R$ feature in Fig. \ref{fig:3}b is a clear signal of such networks.

\par
As I worked on $R$-diagrams in pure fluids with my collaborators, I had occasion to consult Horst about a debate I was having with a journal referee about the Yang-Yang critical point anomaly \cite{Yang1964}.  The referee criticized one of our manuscripts for not mentioning this effect.  Since Horst was visiting me at the time (spring 2012), I asked him about the Yang-Yang anomaly, and it turned out that he knew all about it.  Indeed, he had published measurements on this effect in $^3$He \cite{Brown1972}.  Horst researched the present experimental status of the effect for me, consulting both the literature and his network of ex-students.  Horst then provided me with some recent papers, one a key unpublished experiment.  I ultimately concluded that the experimental evidence for this subtle effect seemed less than overwhelming.  Although this did not help me with my referee, Horst's input was very valuable in allowing me to assess the experimental status of an effect often mentioned in the literature.

\par
During this time period, I also worked on an $R$-diagram for hydrogen, but had to pause on finding that the melting curve for hydrogen was given neither in my primary fluid hydrogen reference, nor in the NIST fluid database \cite{NIST}.  Since Horst had long done research on solid hydrogen, I naturally consulted him about the hydrogen melting curve.  Again, Horst, in communicating with long-time collaborators, was able to supply me with the best recent references for this problem, and I was able to find a fitting formula for the melting curve best matching the range of my fluid data.

\subsection{{\it Black Hole Thermodynamics}}

Geometry of thermodynamics was also applied to black hole thermodynamics, a topic originated by Bekenstein, Hawking, and others in the early 1970s.  The Bekenstein-Hawking formula relates the black hole entropy $S$ to the area $A$ of the event horizon:

\begin{equation} S = \frac{k_B}{4}\left(\frac{c^3}{G\hbar}\right)A,\label{150}\end{equation}

\noindent where $G$, $\hbar$, and $c$ denote, respectively, Newton's gravitation constant, the Planck-Dirac constant $h/2\pi$, and the speed of light.  The idea is to use some theory, perhaps general relativity, to work out $A$ in terms of black hole parameters, and then to use Eq. (\ref{150}) for $S$ to get the full thermodynamics.

\par
The universal proportionality constant between $S$ and $A$ in Eq. (\ref{150}) must contain units of length in order to cancel the length units in $A$.  The only universal constant with units of length is the Planck length:

\begin{equation}\sqrt{\frac{G\hbar}{c^3}}=1.616 \times 10^{-35}\mbox{ m}.\label{160}\end{equation}

\noindent Just by virtue of the physical constants, black hole thermodynamics combines the theories of gravitation, quantum mechanics, and relativity, a reason why there is so much interest in it.

\par
One hope is that analogies with conventional thermodynamics could help in understanding black holes, particularly possible black hole phase transitions.  But making convincing connections between conventional thermodynamic models and black hole models is a difficult game.  For example, black hole thermodynamics includes exotic thermodynamic parameters, such as angular velocity and angular momentum, whose conventional thermodynamic analogs are hard to identify.  Equally challenging is the lack of any consensus regarding the underlying microscopic structure for black hole thermodynamics.  This is a serious deficit since the assessment of conventional phase transitions is always guided by known microscopic structures.

\par
But regardless the thermodynamic scenario, the invariant $R$ always gets calculated by the same methods.\footnote{I got interested in $R$ for black hole thermodynamics when \r{A}man, Bengtsson, and Pidokrajt \cite{Aman2003} worked it out for several simple examples, though not necessarily with my physical interpretation in mind.}  Does the interpretation above in terms of interactions between microscopic constituents extend to the black hole scenario?  In this event, $R$ might tell us something important about a fundamentally hard problem where little is known.  For example, I have observed that at ultralow black hole temperatures (fundamental quasiparticle excitations typically best reveal themselves at very low temperatures) the sign of the black hole $R$ is always positive, with $R\to +\infty$ as $T\to 0$ \cite{Ruppeiner2014}.  This divergence is the same as that for $R$ for a low temperature gas of fermions.  Does this suggest that the fundamental microscopic constituents of black holes are, in fact, fermions?

\subsection{\it{Geometric Equation for the Pure Fluid}}

Let me raise another application where Horst's input was useful to me.  The reader may have noticed that Eqs. (\ref{120}) and (\ref{130}) both contain the correlation volume $\xi^3$.  If $\xi^3$ is eliminated between these two equations, then we get $R$ in terms of $\phi$:

\begin{equation}R=-\frac{\kappa}{\phi},\label{170}\end{equation}

\noindent where $\kappa$ is a dimensionless constant of order unity.  Since $R$ may be written in terms of the derivatives of $\phi$ \cite{Ruppeiner1995}:

\begin{equation} R = \frac{1}{2}\left| \begin{array}{ccc} \phi_{,11}& \phi_{,12}& \phi_{,22}\\ \phi_{,111}&\phi_{,112}&\phi_{,122}\\ \phi_{,112}&\phi_{,122}&\phi_{,222} \end{array}\right| \displaystyle /\left| \begin{array}{cc} \phi_{,11} & \phi_{,12}\\\phi_{,12}&\phi_{,22}\end{array}\right|^2,\label{180}\end{equation}

\noindent the geometric equation Eq. (\ref{170}) is a third-order partial differential equation (PDE) for $\phi$.  Here, I have written the free energy per volume as $\phi=\phi(\beta,h)$, where $\beta=1/k_B T$ and $h=-\beta H$, with $H$ being the ordering field.  The comma notation indicates differentiation $\{1,2\}\leftrightarrow\{\beta,h\}$.

\par
Solving Eq. (\ref{170}) for $\phi$ is expected to be useful in cases with large $\xi^3$, where many molecules have organized into large mesoscopic fluctuating groups.  Scaled, universal thermodynamic properties, independent of the details of intermolecular potentials, are expected here.  It is not unreasonable to think that such properties might emerge from some thermodynamic principle such as the geometric equation.

\par
In the interesting applications so far, the geometric equation was simplified by using a scaled equation of state that reduces the PDE to an ordinary differential equation (ODE), subject to straightforward solution.  This ODE was solved near the critical point \cite{Ruppeiner1991} using the Griffiths' \cite{Griffiths1967} form of the scaled equation of state.  Define the reduced variables

\begin{equation}\{t,m,h\}=\left\{\frac{T-T_c}{T_c},\frac{\rho-\rho_c}{\rho_c},\frac{\mu-\mu(T)}{\mu_c}\right\},\label{}\end{equation}

\noindent where the subscript $''c\,''$ denotes the critical point values, and $\mu(T)$ is the chemical potential on the coexistence curve if $t<0$, and the chemical potential on the critical isochore if $t>0$.  The scaled equation of state is

\begin{equation}\{x,h(x)\}=\left\{t|m|^{-1/\beta},\frac{h}{m|m|^{\delta-1}}\right\},\label{}\end{equation}

\noindent where $\beta$ and $\delta$ are usual critical exponents \cite{Griffiths1967}.  Define also the constants

\begin{equation}\{x_0, h_0\}=\left\{-x|_{h=0,t<0}, h(x)|_{x=0}\right\}.\label{}\end{equation}

\par
\begin{figure}
\centering
\includegraphics[width=3.5in]{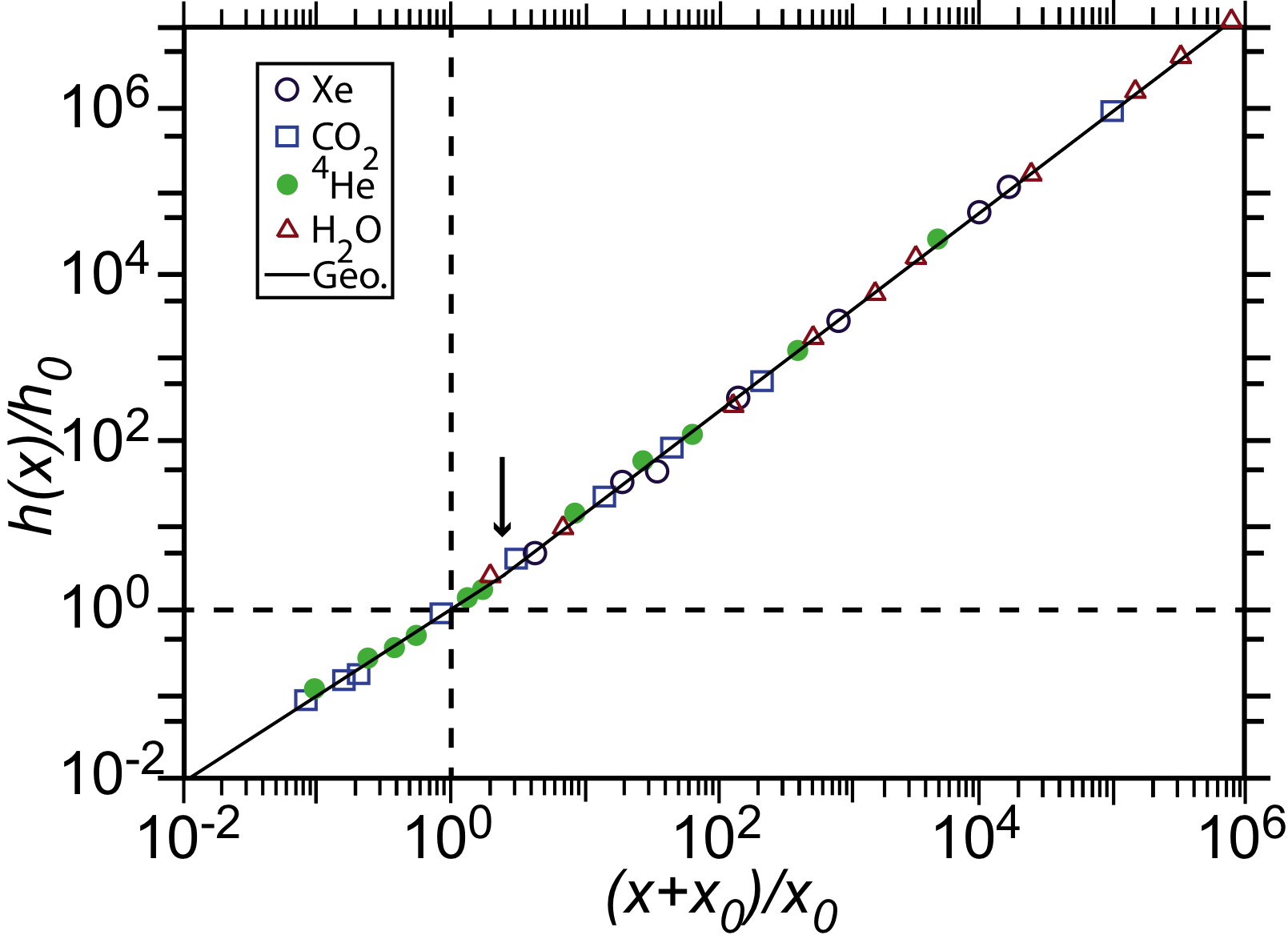}
\caption{The scaled function $h(x)/h_0$ versus $(x+x_0)/x_0$ computed from experiment in four pure fluids and from the theoretical geometric equation.  The down pointing arrow indicates the place where the geometric equation curve suffers a discontinuity in slope.  The match between theory and experiment is excellent.}
\label{fig:4}
\end{figure}

This scaled equation of state is shown in Figure \ref{fig:4}, where I compare experimental data in four fluids with the solution to the geometric equation.  The match between  experiment and theory is excellent.  Computing the theory curve required only two critical exponent values: $\{\beta,\delta\}=\{0.35,4.45\}$.  These values are very near those used in the analysis of the experimental data \cite{Sengers1974}.  The theory curve was computed in two sections, each of which is indistinguishable from a straight line in the graphical display shown in Fig. \ref{fig:4}.  The sections join at the location of the down pointing arrow, which corresponds to a discontinuity in the slope.

\par
The fluid exponent values above differ from the known values of those for the 3D Ising model: $\{\beta,\delta\}=\{0.326, 4.79\}$.  Modern theory puts the critical point for the pure fluid into the same universality class as the critical point for  the 3D Ising model, and so we expect the values of their corresponding critical exponents to be the same.  Measured differences in values are usually attributed to fluid data taken too far from the critical point for the asymptotic properties to fully manifest themselves.  Experimental data supporting this point of view have certainly been collected, including data taken by Horst's group in $^3$He \cite{Pittman1979}.  However, this traditional viewpoint is based on the belief that the discrete lattice gas model properly represents the continuous pure fluid critical regime.  This assumption is perhaps not entirely self-evident.

\par
The discontinuity in the theory slope shown in Fig. \ref{fig:4} is not expected from the pure fluid/3D Ising correspondence since the 3D Ising model is known rigorously to have no discontinuities except along the phase transition curve.  I went through a period (around 1992) of searching old fluid data sets for a jump of about $5\%$ in the isothermal compressibility \cite{Ruppeiner1992}.  I corresponded with Horst about this, and he sent me a detailed tabulation of his group's extensive $^3$He data set \cite{Wallace1970}.  I remember looking at this data, and convincing myself at first that the discontinuity I was looking for indeed existed.  On further reflection, however, I became less certain.  It is just too easy to see discontinuities in discrete and always noisy data, even when none exists.  I never really resolved this point, but I did learn a lot about working with critical point data.

\par
I have noticed recent experimental reports of discontinuous behavior in the supercritical pure fluid regime.  First, along the Widom line extending the phase transition curve in a logical way \cite {Simeoni2010}.  Second, along the Frenkel line close to the melting curve \cite{Bolmatov2015}.  Although neither of these cases corresponds to the change in slope in Fig. \ref{fig:4}, it is nice to see some life to questions of deviations between pure fluids and the 3D Ising model near the critical point.

\subsection{{\it Geometric Equation for the Strongly Interacting Fermi Gas}}

Another area of recent discussion between Horst and myself concerned the ultralow temperature thermodynamic properties of strongly interacting Fermi gases.  This topic is a bit outside of Horst's main research area, but experimental work was done at Duke in diffuse $^6$Li gas by the Thomas group.  I first read about this work in the Duke Physics Department Newsletter.  Strong interatomic interactions created by tuning a magnetic field to Feshbach resonance produce universal thermodynamic properties, the same as for other Fermi systems in this class, such as neutron star material, quark-gluon plasmas, or perhaps even electrons in high-$T_c$ superconductors \cite{Luo2009}.

\par
On reading about the unitary Fermi gas, I thought that I could produce a meaningful solution for it with the geometric equation.  I did some work on this, but it was unclear how I should present my results: as a theory project in the geometry of thermodynamics, or as a data-fitting exercise?  I talked with Horst about this question at the Boston APS March meeting (2012), and he advised me to take the later tack, and to submit to JLTP.  Horst in particular cited the excellent JLTP refereeing.  I took Horst's advice, and was not disappointed.  I ultimately published two papers in JLTP on this topic \cite {Ruppeiner2014u, Ruppeiner2015u}.

\section{Conclusion}

Even old and established theories written up in the textbooks are subject to attempts at improvement.  But for best results, such attempts should be guided by new developments in Physics, and they should be done in a supportive environment where the issues can be worked out and communicated.  Such conditions certainly prevailed at Duke University as I worked in the Horst Meyer group on my ideas on aspects of metric geometry of thermodynamics.

\section{Acknowledgements}

I thank Horst Meyer for his continual support.  While a lot of good work would have been done in geometry of thermodynamics without him on the scene, it probably would not have been done by me.  I also thank my many friends and collaborators.

 \end{document}